\documentclass[12pt,preprint]{aastex}
\usepackage{psfig,lscape}
\def\lap{\lower.5ex\hbox{$\; \buildrel < \over \sim \;$}}
\def\gap{\lower.5ex\hbox{$\; \buildrel > \over \sim \;$}}

\def\ergcm2s{${\rm erg\ cm^{-2}\ s^{-1}}$}

\def\ergscm2s{${\rm erg\ cm^{-2}\  s^{-1}}$}

\def\cm-2{${\rm cm^{-2}}$}
\def\ergs{${\rm erg\ s^{-1}}$}

\begin{document}

\title{Why are X-ray sources in the M31 Bulge so close to Planetary Nebulae?}

\author{Benjamin F. Williams\altaffilmark{1}, Michael~R.~Garcia\altaffilmark{1}, Jeffrey~E.~McClintock\altaffilmark{1}, and Albert~K.~H.~Kong\altaffilmark{1}}
\altaffiltext{1}{Harvard-Smithsonian Center for Astrophysics, 60
Garden Street, Cambridge, MA 02138; \hbox{williams@head.cfa.harvard.edu};
\hbox{garcia@head.cfa.harvard.edu}; \hbox{jem@head.cfa.harvard.edu}; \hbox{akong@head.cfa.harvard.edu}}


\begin{abstract}

We compare a deep (37 ks) {\it Chandra} ACIS-S image of the M31 bulge
to deep [O~III] Local Group Survey data of the same region.  Through
precision image alignment using globular cluster X-ray sources, we are
able to improve constraints on possible optical/X-ray associations
suggested by previous surveys.  Our image registration allows us to
rule out several emission-line objects, previously suggested to be the
optical counterparts of X-ray sources, as true counterparts.  At the
same time, we find six X-ray sources peculiarly close to strong
[O~III] emission-line sources, classified as PNe by previous optical
surveys.  Our study shows that, while the X-rays are not coming from
the same gas as the optical line emission, the chances of these six
X-ray sources lying so close to cataloged PNe is only $\sim$1\%,
suggesting that there is some connection between these [O~III]
emitters (possibly PNe) and the X-ray sources.  We discuss the
possibility that these nebulae are misidentified supernova remnants,
and we rule out the possibility that the X-ray sources are ejected
X-ray binaries.  There is a possibility that some cases involve a PN
and an LMXB that occupy the same undetected star cluster.  Beyond this
unconfirmed possibility, and the statistically unlikely one that the
associations are spatial coincidences, we are unable to explain these
[O~III]/X-ray associations.

\end{abstract}

\section{Introduction}

The high-spatial resolution of the {\it Chandra} X-ray Observatory is
allowing optical counterparts to be found for a large number of
individual extragalactic X-ray sources.  Such a capability provides
the first opportunity to find new examples of exotic sources for which
there are only a handful of Galactic examples.  Because of its
proximity, low extinction, and high mass, M31 is an excellent place to
search for such objects, and associations between X-ray sources and
optical line-emitting sources provide interesting places to start.

Early optical emission-line surveys of M31 (e.g. \citealp{rubin1972})
and X-ray surveys of M31 (e.g. \citealp{vanspeybroeck1979}) produced a
limited number of reliable counterpart candidates due to resolution
and sensitivity issues.  More recently, many digital surveys have
discovered hundreds emission-line sources, including several hundred
compact [O~III] $\lambda5007$ sources in the M31 bulge, believed to be
planetary nebulae (PNe) \citep{ford1978,ciardullo1989}.

At the same time, X-ray spatial resolution has improved dramatically,
with {\it Chandra} providing source locations to a few tenths of an
arcsec. Because the emission-line sources in the M31 bulge are well
separated, even in ground-based images, comparisons of high-resolution
X-ray positions with optical emission-line images of the M31 bulge
reveal interesting counterparts.  Typically these sources of strong
X-ray emission and optical line emission are found to be supernova
remnants (SNRs) \citep{kong2002s,kong2003,williams2004s}, like the
X-ray bright SNRs in our Galaxy.

Several recent surveys of M31 with {\it Chandra}
\citep{kong2002,williams2003c} have revealed possible X-ray
counterparts to optical emission-line sources classified as PNe by
their appearance as point sources in [O~III] images.  Such
counterparts are difficult to understand because PNe are among the
weakest X-ray sources in the Galaxy.  None of the Uhuru or HEAO A-1
X-ray sources was ever optically identified as a PN.  In fact, only in
recent years has it been possible to study Galactic PNe using ROSAT
and {\it Chandra}.  \cite{guerrero2000} detected a total of 13 PNe
with ROSAT; all are faint and have very soft spectra.  {\it Chandra}
observations have provided more detailed information: NGC 6543 and NGC
7293 have peak temperatures (MEKAL model) of 0.5 keV and 1.0 keV, and
X-ray luminosities of $10^{30}$~erg~s$^{-1}$ and 3~$\times
10^{29}$~erg~s$^{-1}$, respectively \citep{guerrero2001}.  The most
luminous planetary we are aware of is NGC 7027 with $L_{x}~\sim~1.3
\times 10^{32}$ erg s$^{-1}$; its temperature is $\sim0.3$~keV
\citep{kastner2001}.

Another possibility is that these PNe are not counterparts at all, but
instead are separated by a few arcsec, close enough to be confused
with counterparts at the accuracy limit of simple catalog
cross-correlation.  To check this possibility, it is necessary to have
images of the PNe sources to register and compare to the X-ray data.
Recently, the Local Group Survey (LGS) team \citep{massey2001} has
released their deep, wide-field emission-line images of M31.  These
images contain X-ray emitting globular clusters (GCs) as well as
previously-cataloged PNe.  These GCs allow precise registration of the
[O~III] and {\it Chandra} images, reducing the X-ray error regions to
a few tenths of an arcsecond.

In this paper, we compare a deep {\it Chandra} ACIS-S image with the
LGS [O~III] image of the M31 bulge in order to tightly constrain the
association between several of the X-ray sources of the M31 bulge and
bright, compact [O~III] emitting regions classified as PNe.  Section 2
describes the data and analysis techniques used.  Section 3 discusses
the comparison of the X-ray positions and the optical emission-line
sources.  Section 4 discusses possible explanations for the
associations, and section 5 provides a summary of our conclusions.

\section{Data}

\subsection{Data Processing}

We obtained the [O~III], [S~II], H$\alpha$, $V$ and $R$-band images of
the M31 bulge from field 5 from the Local Group Survey
(LGS\footnote{http://www.lowell.edu/$\sim$massey/lgsurvey}). These
images have already been properly flat-fielded and the geometric
distortions removed so that the coordinates in the images are good to
$\sim$0.25$''$ on the FK5 system, and the images through different
filters are registered with one another.  We therefore were easily
able to subtract the $V$-band continuum from the [O~III] image in
order to make the [O~III] sources stand out.  The central 40$''$ are
saturated in the continuum image, rendering that section of the data
useless for our purposes.  Fortunately only 2 of the PNe counterpart
candidates lie in this region.  The [S~II] and H$\alpha$ images were
used only to test the flux ratios as a diagnostic for photo-ionization
of the X-ray PNe.

We performed a rough calibration of the LGS [O~III] image by matching
the [O~III] fluxes of 10 planetary nebulae (PNe) with published
[O~III] fluxes \cite{ciardullo1989}.  This calibration provided a
conversion factor of 5.5 $\times 10^{-16}$ erg cm$^{-2}$ ct$^{-1}$.
We also roughly calibrated the H$\alpha$ and [S~II] images by matching
the fluxes of the SNR DDB 1-15 \citep{dodorico1980} to the fluxes
measured in the calibrated data set of \citet{williams1995} (H$\alpha
= 7.3 \times 10^{-14}$ \ergcm2s ; [S~II] = 5.5 $\times 10^{-14}$
\ergcm2s ).  This calibration yielded conversion factors of 1.0
$\times 10^{-16}$ erg cm$^{-2}$ ct$^{-1}$ and 1.8 $\times 10^{-16}$
erg cm$^{-2}$ ct$^{-1}$ in H$\alpha$ and [S II] respectively.  Using
these [O~III], H$\alpha$, and [S II] factors, we converted the LGS
count rates to units of \ergcm2s .

We also obtained a deep {\it Chandra} ACIS-S image centered on the M31
nucleus.  This image had an exposure time of 37.7 ks.  We created
exposure maps for this image using the CIAO script {\it mergeall}, and
we found and measured positions for the sources in the image using the
CIAO task {\it wavdetect}.  We detected 153 sources, 137 of which were
located in regions outside of the saturated portion of the LGS images.
The source list reached a flux limit of $\sim$2.5 $\times$ 10$^{-7}$
ct cm$^{-2}$ s$^{-1}$ (0.3--10 keV), or $\sim$8 $\times 10^{-16}$
\ergcm2s assuming an absorbed power-law spectrum with slope 1.7 and
$N_H = 10^{21}$ cm$^2$, or a (unabsorbed) luminosity limit of $\sim$7
$\times 10^{34}$ \ergs\ in M31, assuming a distance of 780 kpc
\citep{williams2003b}.

The X-ray image contained all but one of the previously mentioned PNe
counterpart candidates \citep{kong2002,williams2003c}, which are all
labeled and marked with their X-ray position error circles in
Figure~\ref{pnims}.  The one counterpart candidate not detected
(r1-23, \citealp{williams2003c}) is an X-ray transient candidate
\citep{kong2002}.  This source is in the saturated central region of
the LGS data; even if it had been detected in the ACIS-S data, its
position relative to the PNe could not have been further constrained
by this study.

\subsection{Image Alignment}

We aligned the coordinate system of the ACIS-S image with the LGS
coordinate system by translating and adjusting the plate scale of the
ACIS-S coordinate system so that 13 globular cluster sources had the
same coordinates as the centroids of the respective globular clusters
in the LGS $V$-band image.  This transformation, performed using the
IRAF\footnote{IRAF is distributed by the National Optical Astronomy
Observatory, which is operated by the Association of Universities for
Research in Astronomy, Inc., under cooperative agreement with the
National Science Foundation.} task {\it ccmap}, had root-mean-square
residuals of 0.16$''$ in RA and 0.15$''$ in DEC.  These errors were
added in quadrature to the position errors of the sources, determined
by {\it wavdetect}, to calculate the final position errors for X-ray
sources on the LGS [O III] image.

\begin{figure}
\centerline{\psfig{file=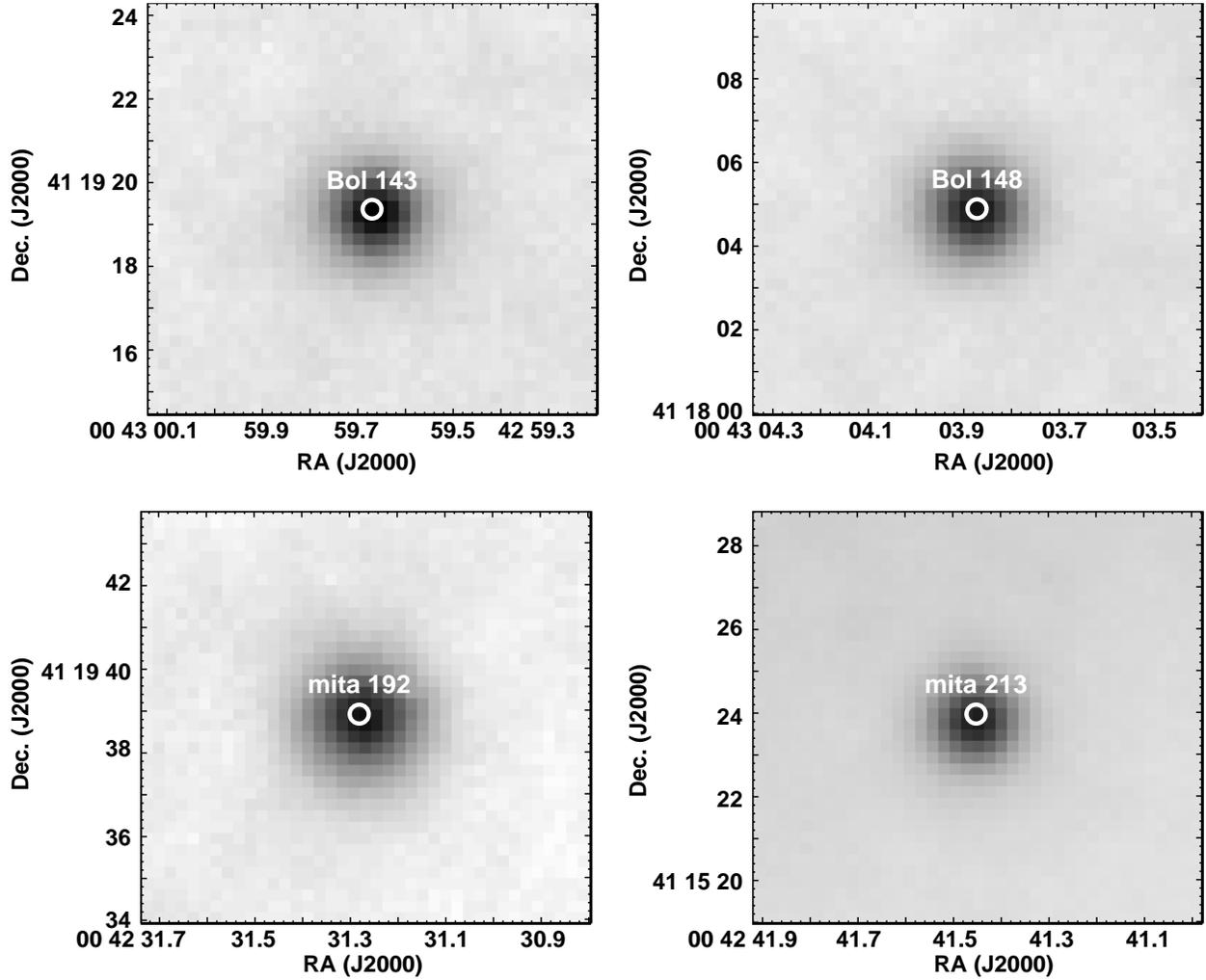,width=6.5in,angle=0}}
\caption{Four sections of the LGS [O~III] image are shown.  The
1$\sigma$ error circles of the positions for strong X-ray sources
located in 4 globular clusters are overplotted.  The X-ray positions
are clearly within the errors of the globular cluster centers,
revealing the precision of the X-ray/optical alignment.}
\label{gcims}
\end{figure}

Figure~\ref{gcims} shows the error circles of four of the globular
cluster X-ray sources used to align the X-ray and optical images
plotted on the [O III] image.  The quality of the alignment is within
the position errors of the globular cluster sources.  The globular
cluster X-ray sources in the aligned images are within 1$\sigma$ of
the centers of the optical globular clusters.  On the other hand, the
PN sources, shown in Figure~\ref{pnims}, are clearly not within the
calculated errors of the X-ray sources.  These slight offsets between
the X-ray sources and the [O~III] nebulae show that the [O~III]
emission is not coming from the same source as the X-ray emission, but
the proximity of the sources in so many cases is intriguing, as it
suggests the sources are connected in some way.

\subsection{X-ray Spectra}

To further investigate the nature of these peculiar X-ray sources, we
extracted their X-ray spectra from the ACIS-S observation using the
CIAO task {\it psextract}.  We binned the spectra so that they
contained $\gap$10 counts per bin, allowing standard $\chi^2$
statistics.  Source r1-2 contained sufficient counts to provide
$\gap$20 counts per bin.  

We then fit the background-subtracted spectra from 0.35--7 keV using
the CIAO 3.0/Sherpa package \citep{freeman2001}.  We tried fits using
Raymond-Smith (RS), blackbody, and power-law models, including
absorption and a correction of the instrumental response for the
well-known contamination build-up on the ACIS detectors.  The results
of the fits are listed in Table~\ref{spectab} and discussed below in
\S 3.2.

\section{Results}

\subsection{X-ray source locations}

\begin{figure}
\centerline{\psfig{file=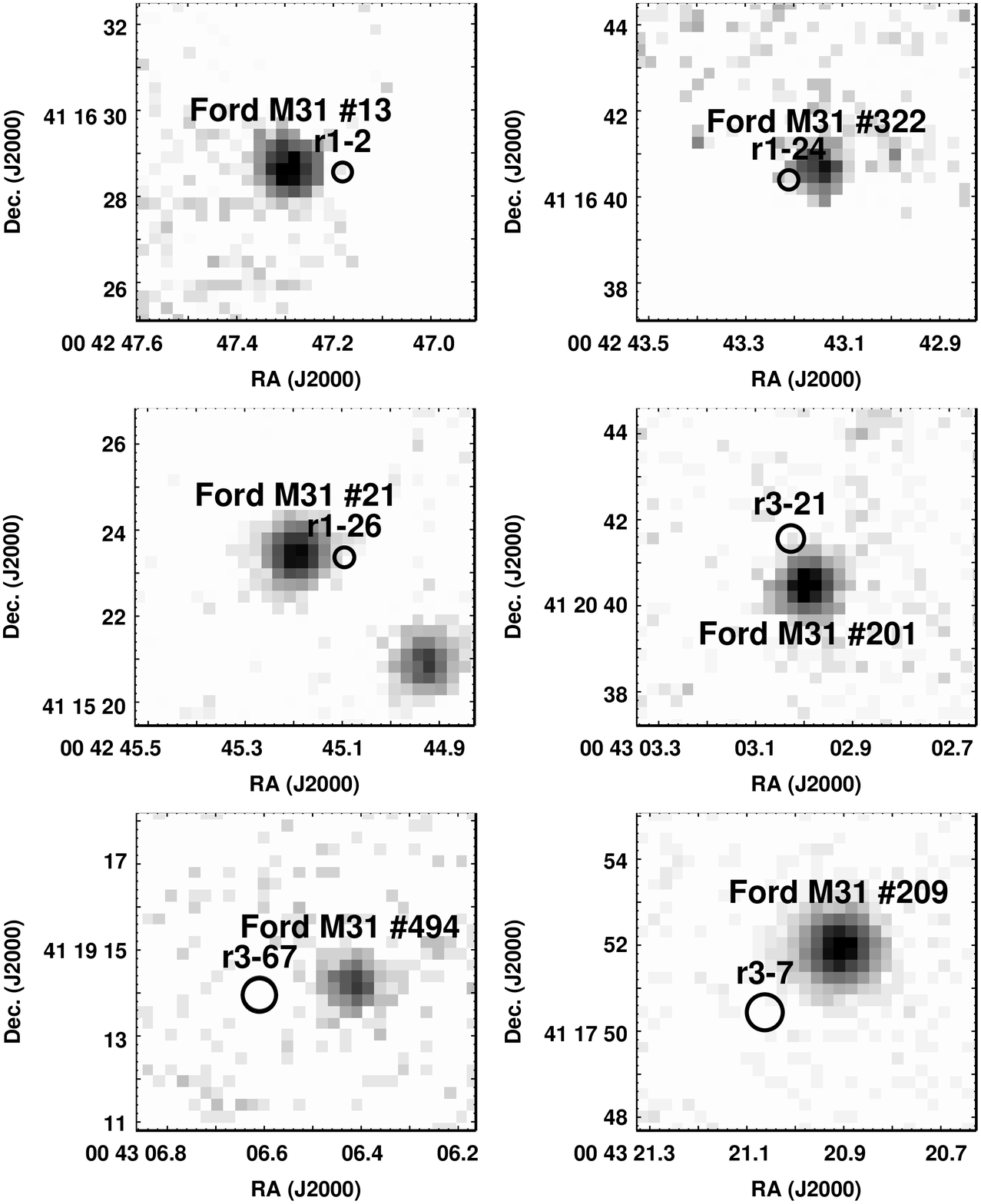,height=7in,angle=0}}
\caption{The error circles of {\it Chandra} X-ray positions of sources
associated with [O~III] nebulae classified as PNe in M31.  The
1$\sigma$ error circles are plotted on sections of an aligned,
continuum-subtracted [O~III] image.  The [O~III] nebulae show up as
dark areas on the images, and they are labeled with their PN catalog
names. }
\label{pnims}
\end{figure}

We visually searched the aligned [O~III] image for nearby X-ray
sources by placing 2.5$''$ radius circles onto the [O~III] image
centered on the locations of all X-ray sources detected in the ACIS-S
images.  This search yielded 15 counterpart candidates, 6 of which
have been cataloged as PNe.  Figure~\ref{pnims} shows these 6 ``PNe''
sources within 2.5$''$ of the X-ray source position.  The 1$\sigma$
error circles for the X-ray positions are plotted on the images.  

Five of the other 9 nearby [O~III] sources are cataloged SNRs (r2-56,
r2-57, r3-63, r3-69, and r3-84
\citep{kong2002s,kong2003,williams2004s}.  The [O~III] sources near
CXOM31 J004229.1+412857 \citep{kaaret2002} and r3-87 (CXOM31
J004226.1+412552, \citealp{kong2002}) are uncatalogued.  The [O~III]
source near CXOM31 J004220.5+412640 \citep{kaaret2002} is a known
emission line source (WB92a 26, \citealp{walterbos1992}) and radio
source ([B90] 64, \citealp{braun1990}). The source near r1-16 is a
cataloged emission line object of unknown nature \citep{wirth1985}.

While all of these [O~III]/X-ray associations are of interest, the
SNRs have been studied in detail in previous publications, and the
other sources have yet to be classified.  Herein we focus on the 6
X-ray sources within 2.5$''$ of ``PNe''.  Of these objects, r3-67 is
also within 2.5$''$ of a known radio source ([B90] 122,
\citealp{braun1990}).  The radio source is more closely aligned with
the X-ray source (1.7$''$) than the [O~III] nebula (3.7$''$).

In order to determine the number of random X-ray/optical associations
that would be found by searching within 2.5$''$ of detected X-ray
sources, we shifted the positions of our X-ray sources by 15$''$ and
repeated the visual search for counterparts.  This experiment was
performed 4 times in 4 directions of position shifts.  The mean number
of random association found was 3, with an average number of cataloged
PNe associations of 2.  We therefore expect that 2 of our associations
with PNe are chance superpositions; however, this number still
suggests that 4 of the X-ray sources are somehow associated with PNe.
According to Poisson statistics, there is only a 1.2\% chance that all
6 of these associations are chance superpositions.  \footnote{A
2.5$''$ search radius was chosen as the root-sum-square of the typical
position errors in the {\it Chandra} and Ford catalogs.  It should be
noted that if we run the same test with a 1.5$''$ search radius, we
find 4 PNe associations and an average of 1 random association.  These
numbers leave a 1.5\% chance that all 4 of the associations within
1.5$''$ are chance superpositions, suggesting that the statistics are
not strongly dependent on the exact search radius chosen.}

For a more sophisticated determination of the likelihood of 6 false
PN/X-ray associations, we performed monte carlo tests for chance
associations as a function of the separation between the PN and X-ray
source.  Stepping from maximum separation angles of 1$''$ to 3$''$
in 0.1$''$ increments, we searched for PN/X-ray associations by
comparing the X-ray and PN catalogs after rotating them about the M31
nucleus with respect to one another by a random angle (20--340
degrees).  Any associations found in the rotated catalogs could not be
real; therefore these searches provided the number of expected false
associations for each maximum separation angle tested.

\begin{figure}
\centerline{\psfig{file=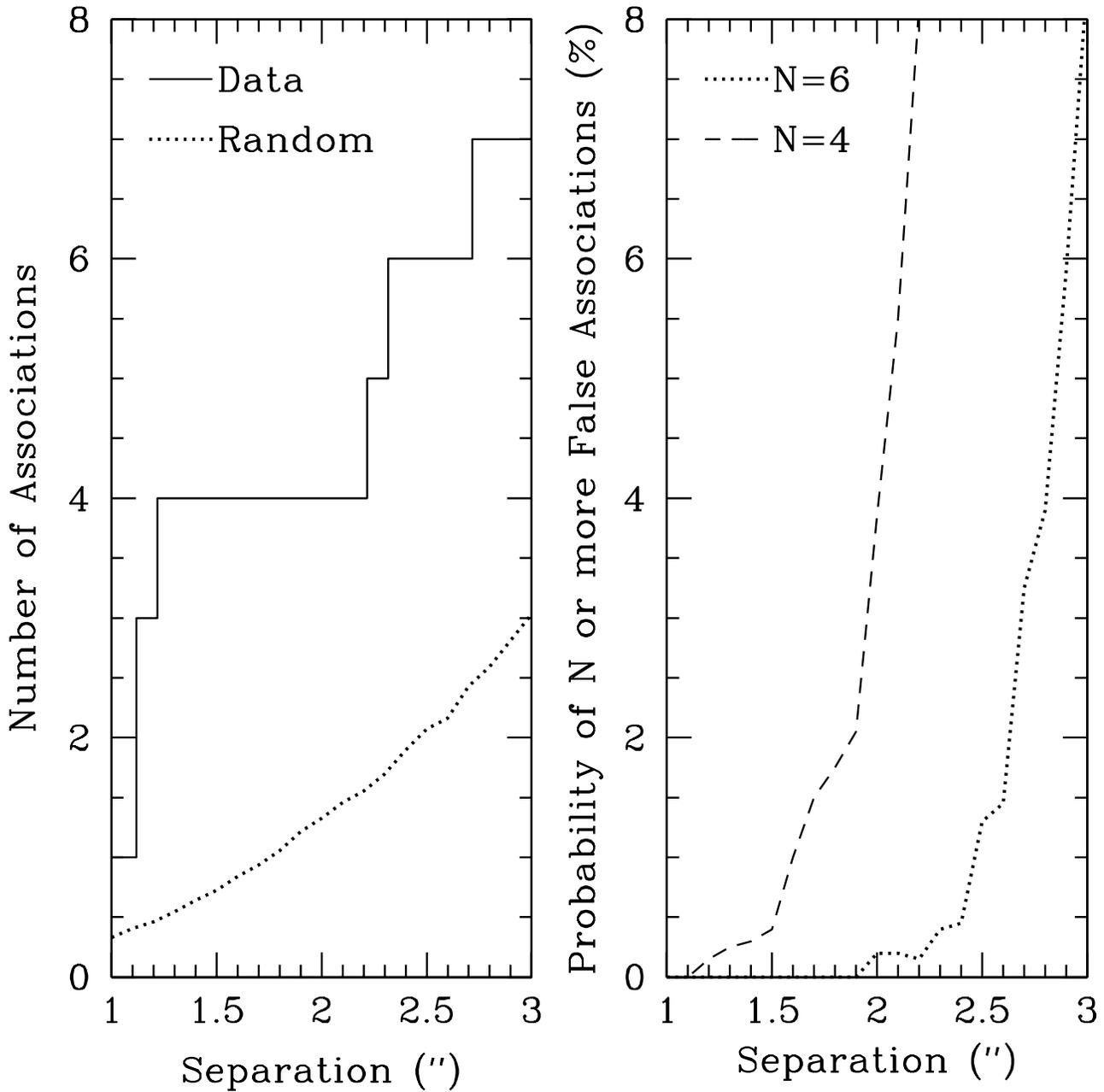,height=7in,angle=0}}
\caption{The results of our monte carlo tests for coincidental matches
of PNs and X-ray sources as a function of maximum separation. {\it
Left Panel:} The histogram shows the number of matches found in the
real, aligned data.  The dotted line shows the mean number of false
associations. {\it Right Panel:} The dotted line shows the probability
of 6 or more false associations in the randomly rotated data, and the
dashed line shows the probability of 4 or more false associations.}
\label{mc}
\end{figure}

We performed the monte carlo test using 2000 random rotation angles
for each maximum separation angle.  This number of trials provided
a robust mean number of false associations, shown in Figure~\ref{mc}.
In addition we monitored the percentage of tests that produced 4 or
more false associations as well as the percentage that produced 6 or
more false associations.  These results are also shown in
Figure~\ref{mc}.

Our monte carlo tests confirmed what we found with our searches by
eye, finding a mean of 2.07 false associations within 2.5$''$
separation and 0.73 false associations within 1.5$''$ separation.  The
monte carlo results also support our use of Poisson statistics for
estimating the likelihood of false associations, as 1.3\% of tests
find 6 false associations within 2.5$''$ and 0.4\% find 4 false
associations within 1.5$''$.  Furthermore, as shown in Figure~\ref{mc}
we find 6 associations at a range of maximum separation angles
(2.3$''$--2.7$''$).  Over this range, the probability of finding 6
false associations changes rather steeply (see Figure~\ref{mc}),
covering a range of 0.4\%--3.25\%.  This probability range represents
our most conservative measurement of the probability that all of these
associations are false.

Finally, our precise image alignment provides position errors for the
X-ray source locations on the LGS [O~III] image of only a few tenths
of an arcsec.  These small position error circles are shown in
Figure~\ref{pnims}.  All of our X-ray sources are slightly, but
significantly, offset from the PNe positions.  This offset is
intriguing.  It suggests that, while the X-ray emission is not likely
to be coming from the PNe itself, the X-ray sources are likely somehow
associated with the PNe.

\subsection{X-ray properties}

We analyzed the spectra of all six X-ray sources that lie within 2.5$''$
of PNe.  The spectral parameters for the absorbed power-law fits,
including model type, model parameters with 1$\sigma$ errors,
$\chi^2/{\nu}$, and absorption-corrected 0.3-7 keV luminosities, are
given in Table~\ref{spectab}.  In one case, r3-7, the best fit
absorption fell below the known foreground absorption.  We therefore
accepted the best fit model with the absorption fixed at the
foreground value of 6$\times$10$^{20}$ cm$^{-2}$
(e.g. \citealp{kaaret2002,trudolyubov2002a}).

In one case, an absorbed RS model provided a fit as good as the fit
from an absorbed power-law.  Source r3-21 is as well fit by an
absorbed RS model with $N_H$ = 9$\pm$3 $\times 10^{21}$ cm$^{-2}$ and
kT = 4$\pm$2 keV.  Even in this case, the absorbed power-law model is
as likely as the RS model to be the correct interpretation of the
data.  Therefore, Table~\ref{spectab} shows the power-law fit, as it
does for the other 5 sources.

In the other 5 cases, an absorbed power-law model provided the only
fits with $\chi^2/\nu < 1.5$.  In three of these cases, the absorbed
power-law fit was improved by the addition of a blackbody component to
the spectrum.  We applied an F-test to the $\chi^2$ values of the
spectral fits with the addition of a blackbody component
(e.g. \citealp{bevington}).  
The $F_{\chi}$ values for r1-26, r1-2, and r3-67 were 23.1, 15.6, and
21.1, respectively (the others had $F_{\chi} < 1$).  These values
leave $<$1\% probability that a blackbody component should not be
included in the fits for r1-26 and r1-2, and $<$5\% probability that
the component should not be included in the fit to r3-67.  The
blackbody components comprise 68\%, 24\%, and 99\% of the modeled
unabsorbed luminosities (0.3--7 keV) of r1-26, r1-2, and r3-67,
respectively.

The sources all show somewhat similar spectral properties.  For
example, all have power-law components with slopes consistent with
1.8--2.0, and all but r3-21 are consistent with $N_H \leq 2 \times
10^{21}$ cm$^{-2}$.  Even the absorption column of r3-21 is within
1.8$\sigma$ of this value.

The X-ray variability of these sources has been investigated in
\citet{kong2002}.  Object r1-24, r1-2, and r3-7 were found to have
varying intensity in X-rays.  The other 3 sources have constant X-ray
flux.  These properties allow that these sources could be XRBs.

\subsection{Optical properties}

We estimated the narrow-band fluxes of these sources in the LGS
images.  These fluxes were measured in 2.5$''$ diameter apertures
centered on the position of the [O III] nebula as determined by the
IRAF task {\it imcentroid}.  The counts in each nebula in each band
were measured using the IRAF task {\it phot} and then converted to
flux units with the calibration discussed in \S 2.  The resulting
fluxes are provided in Table~\ref{nbtab}.  

The [S~II]/H$\alpha$ ratio of line-emitting regions is a well-known
diagnostic of shock heating, where ratios $>$0.4 are typical of shock
heated regions, such as SNRs, and ratios $<$0.4 are typical of
photo-ionized regions, such as H II regions or PNe
\citep{levenson1995}.  Ford~322, Ford~21, Ford~13, and Ford~201 have
[S~II]/H$\alpha$ ratios of 0.9, 1.4, 1.1, and 0.7 respectively,
suggesting that these nebulae are shock-heated, and not photo-ionized.
We are not able to determine the [S~II]/H$\alpha$ ratio for Ford 494
due to lack of a [S~II] or H$\alpha$ detection.  Only for Ford 209 is
the [S~II]/H$\alpha$ ratio consistent with a PNe hypothesis.  

The [O~III]/H$\alpha$ ratios (see Table~\ref{nbtab}) were corrected
for absorption because of the large difference in wavelength.  We
estimated the extinction using the relationship of \cite{predehl1995}.
We assumed a standard reddening law, and we assumed $A_V \sim
A_{[O~III]}$ and $A_R \sim A_{H\alpha}$. We compared our [O~III]
fluxes and [O~III]/H$\alpha$ ratios to the objects in common with
Ciardullo et al. (2002).  Our [O~III] flux estimates for Ford 13 and
Ford 21 are in agreement with their measurements, as is our estimate
of the [O~III]/H$\alpha$ ratio for Ford 13.

On the other hand, our [O~III]/H$\alpha$ ratio estimate for Ford 21 is
larger than theirs by a factor of 3.  This suggests that our H$\alpha$
flux estimate could be significantly low, causing the very high
[S~II]/H$\alpha$ ratio (1.4). This discrepancy underscores the
difficulty of measuring faint source fluxes in the high background of
the M31 bulge.  Ford 21 is the faintest PN we detect in H$\alpha$ (1.1
$\times$ 10$^{-15}$ erg s$^{-1}$) in an area of very high background
(background/source = 112).  For comparison, Ford 209, the faintest PN
in our sample in H$\alpha$ (0.9 $\times$ 10$^{-15}$ erg s$^{-1}$), has
a background/source flux ratio of 13.  The high background of Ford 21
hinders the measurement of precise fluxes.  We note that, although our
flux ratio estimates are not precise enough to yield reliable source
classifications, even if the H$\alpha$ flux is truly a factor of 3
stronger than our estimate, the [S~II]/H$\alpha$ ratio is $\sim$0.5,
still high for a PN.

[O~III]/H$\alpha$ ratios are not reliable for indicating shock-heating
or photo-ionization, as photo-ionized regions cover a wide range of
[O~III] emission strengths \citep{ho1993,veilleux1987,mccall1985};
however, it is interesting to note that four of the [O~III]/H$\alpha$
ratios (Ford 322, Ford 13, Ford 201 and Ford 209) are $<$5.  These
values are not out of the ordinary for interstellar shocks containing
a variety of shock velocities $\gap$100 km~s$^{-1}$
(e.g. \citealp{vancura1992,fesen1997,mavromatakis2000}).  The other
two nebulae have [O~III]/H$\alpha$ ratios $>$10, suggesting
photo-ionization, very high shock velocities, or O-rich emission
regions.  If these sources are SNRs, they could be young with high
shock velocities: a characteristic consistent with their small sizes.

While full optical spectra will be required to determine the true
nature of these sources, the suggestion of shock heating from the
narrow-band flux ratios provides further evidence that 4 or 5 of these
6 objects are not PNe.

\section{Discussion}

Because there is very little likelihood that these 6 X-ray sources lie
so close to classified PNe by chance, we assume for the moment that at
least some of these sources are somehow associated with the nearby
[O~III] nebulae.  It should be noted that, even if the positions were
coincident, there is no current model for, or example of, PNe that
produces strong X-ray emission.  There are also no models for X-ray
binaries that produce strong [O~III] emission.  The unique Galactic
X-ray source GX1+4 emits in [O~III]; however, even this extreme
example has [O~III]/H$\alpha \ll 1$ \citep{chakrabarty1997}.

If these are true X-ray/[O~III] associations, what are they?  The
spectral indices and absorption indicated for the X-ray sources are
typical of X-ray binaries.  However, there are no PN formation models
that would predict the presence of a nearby XRB.  If these [O~III]
nebulae are SNRs, the XRBs may have been formed in the same supernova
event, but we argue below that even this scenario is unlikely.

\subsection{Are the [O~III] Nebulae SNRs?}

Before we address the nature of the X-ray sources, we attempt to
understand the nature of the nearby [O~III] nebulae.  Their
[S~II]/H$\alpha$ ratios and proximity to strong X-ray sources make
these nebulae very strange indeed, as they are unresolved, strong
[O~III] sources in M31 that are likely shock-heated.  In addition, any
hypothetical X-ray sources coincident with the [O~III] nebulae are not
detected by {\it Chandra}, requiring that they are fainter than
$\sim$10$^{35}$ \ergs\ in X-rays (see \S 2.1).

There is a possibility, as several of the [S~II]/H$\alpha$ flux ratios
suggest, that the [O~III] nebulae are SNRs.  If these [O~III] nebulae
are SNRs, they must be young.  To be unresolved in the 1$''$ seeing of
the LGS images, these SNRs would need to be $\lap$4 pc across, smaller
than the sizes of SN 1006 ($\sim$7 pc, \citealp{willingdale1996};
$\sim$19 pc, \citealp{winkler2003}) and 3C~58, thought to be the
remnant of the supernova of 1181 C.E. ($\sim$5 pc,
\citealp{vandenbergh1990}).  These comparisons limit the ages of the
SNRs to $\lap$1000 yr.

The low X-ray luminosities ($\lap$10$^{35}$ \ergs ) do not rule out
the possibility that these nebulae are SNRs, despite their implied
ages.  While the recent X-ray detections of several SNRs in M31
\citep{kong2003,williams2004s} show that some SNRs are located in
regions of the ISM with densities $>$0.1 cm$^{-3}$,
\cite{magnier1997x} found that most SNRs in M31 are weak in X-rays,
suggesting that the density of the ISM is lower in the vicinity of
many M31 SNRs.

Furthermore, young SNRs of both Type Ia and Type II can be weak in
X-rays.  For example, SN 1006 and 3C~58 are thought to have been Type
Ia \citep{winkler2003} and Type II \citep{panagia1980} supernovae,
respectively. SN~1006 has an X-ray luminosity (1--10 keV) of $\sim$5
$\times 10^{34}$ \ergs \citep{winkler1976,winkler2003}, and 3C~58 has
an X-ray luminosity (0.1--4 keV) of only $\sim$10$^{34}$ \ergs
\citep{helfand1995}.  Therefore neither of these young SNRs would have
been detected by the {\it Chandra} surveys.

Finally, if the [O~III] nebulae are SNRs, some of them must have old
progenitors in order to explain their location in the M31 bulge.
Three of these nebulae (Ford 13, 21, 322) lie within a 2$' \times 2'$
box centered on the nucleus, in the old stellar population of the M31
bulge.  Although the M31 bulge is known to contain some molecular
clouds \citep{melchoir2000} and SNRs
\citep{sjouwerman2001,kong2003,williams2004s}, allowing the
possibility of some recent star formation, it is not known to harbor a
young stellar population \citep{deharveng1982,stephens2003}. If these
[O~III] nebulae are SNRs, their progenitor stars were likely old.

Because the M31 bulge does not contain a substantial number of young,
high-mass stars, if these ``PNe'' are in fact SNRs, we must consider
the possibility that some are SNRs from Type Ia supernovae.  Such
supernovae likely have old, white dwarf progenitors
(e.g. \citealp{hoeflich1998}).  Type Ia supernovae do not arise from
the core-collapse of a massive star; they do not produce stellar-mass,
compact objects (i.e. XRBs).  Even if these inner three [O~III]
nebulae are SNRs, the fact that the progenitor supernovae did not
likely produce XRBs provides the first hint that these [O~III]/X-ray
associations cannot be SNR/XRB pairs.

\subsection{Are the X-ray Sources Ejected XRBs?}

Returning to the six strong {\it Chandra} X-ray sources displaced from
the [O~III] nebulae, we address the question of how these sources
could be related to the neighboring nebulae.  The X-ray sources
certainly cannot be related to the nebulae if the nebulae are PNe, but
if the nebulae are SNRs, the X-ray sources might be ejected X-ray
binaries.  To check this hypothesis, we assume that the supernovae
were able to create XRBs, temporarily ignoring the possibility that,
if the nebulae are SNRs, some of them had Type Ia progenitors.

The idea that these slightly misaligned sources are ejected binaries
is not entirely radical.  Such an association has been suggested in
the Galaxy for the X-ray binary Cir X-1.  While the initial suggestion
of an association with the nearby SNR G321.9-0.3 has been ruled out
\citep{mignani2002}, there is a new suggestion that the radio
synchrotron nebula which surrounds Cir X-1 is actually its birth SNR
\citep{clarkson2004}.

If these [O~III] nebulae are SNRs, the comparisons of their sizes to
SN~1006 and 3C~58 (see \S 4.1) limit the ages of the SNRs to
$\lap$1000 yr.  If these SNRs produced the nearby XRBs, and the XRBs
are 2.8--8.7 pc away from the SNRs' centers, these XRBs must be moving
at least 2700--8500 km~s$^{-1}$.  Such velocities would be higher than
those observed for Galactic neutron stars, which typically range from
100--1000 km~s$^{-1}$ \citep{cordes1998}.

Single neutron stars with very high velocities ($\gap$3000 km
s$^{-1}$) may conceivably exist.  \cite{arzoumanian2002} suggest that
15\% of neutron stars have velocities $>$1000 km~s$^{-1}$ and 10\% of
neutron stars younger than 20 kyr will lie outside of their birth SNR.
In addition, selection effects in Galactic pulsar surveys could lead
to an underestimate of the number of high-velocity neutron stars
\citep{cordes1998}.  Although such claims favor the existence of
high-velocity neutron stars, there is no evidence for neutron star
velocities $>$3000 km s$^{-1}$.

Moreover, it is difficult to imagine how a binary system could be born
with a velocity of $>$3000 km~s$^{-1}$.  Any pre-supernova binary
would have an orbital velocity that is only a small fraction of 3000
km~s$^{-1}$, and the asymmetric kick received by the compact object
during the supernova event would be so fierce that it would unbind the
binary.  We therefore conclude that, whether the [O~III] nebulae are
SNRs or not, it is unlikely that even one of these six X-ray sources
could be a high-velocity X-ray binary ejected from a SNR.

\subsection{Do the X-ray/PN Pairs Reside in Star Clusters?}

There is an additional non-random way in which a PN and an X-ray
source could be so close to one another.  The PN and XRB could be in
the same star cluster, putting them spatially near one another while
being unrelated physically.  In such a scenario, the progenitors of
the XRB and PN must have been formed at the same time.  The neutron
star in the XRB would have then formed from its high-mass progenitor
before the PN formed from its lower mass progenitor.  Applying this
possibility to the M31 associations requires that the star clusters
themselves have gone undetected.

There are Galactic examples of LMXB/PN associations.  We
cross-correlated the Galactic PN catalog of \citet{kohoutek2001} with
the Galactic LMXB catalog of \citet{ritter2003}.  This
cross-correlation revealed only two associations with separations less
than 250$''$ (at typical Galactic distances 250$''$ corresponds to
$\sim$2.5$''$ at the distance of M31).  Both of these associations
involve a PN and an LMXB that lie in the same globular cluster.  The
PN PK~065-27~1 and the LMXB 4U~2129+12 are 30$''$ apart in NGC~7078
(M~15).  The PN JaFu~2 and the LMXB 4U~1746-37 are 30$''$ apart in
NGC~6441.  There is a second PN inside the tidal radius of NGC~6441
(7$'$; \citealp{bahcall1976}); PN H 1-36 has a separation from
4U~1746-37 of 310$''$.  Monte carlo tests, in which we rotated the
Galactic catalogs with respect to one another by random angles (0.1--5
degrees), show that in 1000 iterations there is $<$0.1\% probability
of 2 chance associations with separation less $<$100$''$ and 5.5\%
probability of 2 chance associations with separation $<$250$''$.

Apparently, PNe and LMXBs that are close to one another in the Galaxy
occupy the same globular cluster.  However, globular clusters like
NGC~7078 and NGC~6441 are easily detected in ground based images of
M31.  For example, NGC~6441 would have $V\approx17.5$ if located in
M31, and would be easily detected like other globular clusters within
1$'$ of the M31 nucleus with $V\gap18$ (e.g. ACH~6;
\citealp{auriere1992}).  No such clusters are detected at the
locations of any of the [O~III]/X-ray associations.

Clearly if these M31 LMXB/PN associations are located in clusters of
stars, they are not globular clusters. In recent years, the
distinction between globular clusters and open clusters has been
blurred.  Studies of extragalactic star clusters are revealing that
they may form according to a power-law mass function (e.g. the
Antennae; \citealp{zhang1999}).  Each cluster then stays bound for a
length of time determined by its mass, concentration, and orbital
parameters \citep{fall2001}.  The fact that the star clusters in the
Galaxy happen to fall neatly into two categories (open and globular)
is due to the cluster formation history of the Galaxy.  Other galaxies
have different cluster formation histories and are likely to contain
some star clusters that do not fit into these two categories.
Examples of different cluster types in local galaxies include the
``blue globular'' clusters in the Large Magellanic Cloud
\citep{hodge1984} and the massive and compact young star clusters in
M31 \citep{williams2001a}.

The star clusters hosting the LMXB/PN associations in M31 would need
to be the diffuse, faint relics of once-larger clusters.  However, the
likelihood that such relic clusters would be massive enough to contain
both an LMXB and a PN simultaneously, massive enough not to dissociate
completely in $\sim$10$^9$ years, and faint enough to be undetected,
seems low.

\section{Conclusions}

Several recent {\it Chandra} surveys of M31 have identified X-ray
sources in the M31 bulge spatially coincident with previously
cataloged PNe.  Using the [O~III] data from the Local Group Survey, we
have directly compared the X-ray source positions to the locations of
the cataloged PNe, finding that the X-ray sources are not precisely
coincident with the PNe.  The estimated [S II]/H$\alpha$ ratios of 4
of these ``PNe'' are consistent with shock-heating, i.e. SNRs.

The fact that the X-ray emission is displaced from the [O~III]
emission is consistent with the classification of these nebulae as
PNe, which are notoriously feeble X-ray sources.  At the same time,
the spatial proximity is unlikely to be coincidental: there is only a
0.4\%--3.25\% probability of these associations occurring by chance.
These associations therefore require some explanation.  Any
explanation that physically links these strong X-ray sources to the
[O~III] sources does not allow the [O~III] sources' PNe
classifications to stand.

Only detailed spectroscopy is likely to determine the true nature of
these [O~III] sources and the reason for their proximity to bright
X-ray emitters.  If (as our narrow-band flux estimates suggest) they
are X-ray weak, young SNRs like SN~1006 or 3C~58, the existence of the
nearby X-ray sources is still unexplained.  The nearby X-ray sources
cannot be ejected XRBs.  As stated above (see \S 4), such a scenario
requires core-collapse supernovae and supernova kick velocities
$\gap$3000 km s$^{-1}$ to explain the distances from the nebulae to
the X-ray sources.  These kick velocities would disrupt any X-ray
binary system.

Finally, perhaps an XRB/PN association could be produced within a
relic of a star cluster.  This explanation requires underlying star
cluster relics in M31 massive enough not to dissociate completely in
$\sim$10$^9$ years and faint enough to be undetected in ground-based
images.  Otherwise, the best explanation for these associations
appears to be the statistically unlikely one that all of these
[O~III]/X-ray pairs are spatial coincidences.


We thank John Raymond for advising in the interpretation of the
[O~III] fluxes.  Support for this work was provided by NASA through
grant number GO-9087 from the Space Telescope Science Institute and
through grant number GO-3103X from the {\it Chandra} X-ray Center.
MRG acknowledges support from NASA LTSA grant NAG5-10889.

\vspace{-0.5cm}

\clearpage

\begin{landscape}
\footnotesize
\begin{deluxetable}{cccccccccccc}
\footnotesize
\tablewidth{8.2in}
\tablecaption{X-ray sources near cataloged PNe}
\tableheadfrac{0.01}
\tablehead{
\colhead{\footnotesize{OBJ}} &
\colhead{\footnotesize{RA}} &
\colhead{\footnotesize{Dec.}} &
\colhead{\footnotesize{PN}} &
\colhead{\footnotesize{Sep ($''$)}} &
\colhead{\footnotesize{Slope}} & 
\colhead{\footnotesize{kT (keV)\tablenotemark{a}}} &
\colhead{\footnotesize{$N_H$\tablenotemark{b}}} &
\colhead{\footnotesize{$\chi^2/\nu$}} &
\colhead{\footnotesize{$Q$\tablenotemark{c}}} &
\colhead{\footnotesize{\#cts}} &
\colhead{\footnotesize{$L_X$\tablenotemark{d}}}
}
\tablenotetext{a}{For sources with a significant blackbody component, the blackbody temperature is provided with a 1$\sigma$ error.}
\tablenotetext{b}{The absorption with 1$\sigma$ error in units of $10^{21}$ cm$^{-2}$.}
\tablenotetext{c}{The probability (based on $\chi^2/\nu$) that the data represent a sample taken from a source spectrum with the model parameters listed.}
\tablenotetext{d}{The unabsorbed X-ray luminosity (0.3--7 keV) in units of $10^{36}$ erg s$^{-1}$, assuming the spectral fit represents the true source spectrum.}
\startdata
\footnotesize{r1-24} & \footnotesize{00:42:43.211} & \footnotesize{+41:16:40.39} & \footnotesize{Ford 322} & \footnotesize{0.74} & \footnotesize{2.1$\pm$0.2} & \nodata & \footnotesize{2.5$\pm$0.7} & \footnotesize{17.29/19} & \footnotesize{0.570} & \footnotesize{227} & \footnotesize{4.4}\\
\footnotesize{r1-26} & \footnotesize{00:42:45.095} & \footnotesize{+41:15:23.36} & \footnotesize{Ford 21} & \footnotesize{1.11} & \footnotesize{1.79$\pm$0.03} & \footnotesize{0.09$\pm$0.02} & \footnotesize{2$\pm$1} & \footnotesize{5.77/13} & \footnotesize{0.954} & \footnotesize{194} & \footnotesize{6.3}\\
\footnotesize{r1-2} & \footnotesize{00:42:47.182} & \footnotesize{+41:16:28.56} & \footnotesize{Ford 13} & \footnotesize{1.24} & \footnotesize{1.88$\pm$0.02} & \footnotesize{0.84$\pm$0.05} & \footnotesize{1.5$\pm$0.1} & \footnotesize{142.5/127} & \footnotesize{0.164} & \footnotesize{3599} & \footnotesize{63.8}\\
\footnotesize{r3-21} & \footnotesize{00:43:03.027} & \footnotesize{+41:20:41.56} & \footnotesize{Ford 201} & \footnotesize{1.17} & \footnotesize{2.0$\pm$0.5} & \nodata & \footnotesize{9$\pm$4} & \footnotesize{8.63/6} & \footnotesize{0.195} & \footnotesize{93} & \footnotesize{4.0}\\
\footnotesize{r3-67} & \footnotesize{00:43:06.610} & \footnotesize{+41:19:13.98} & \footnotesize{Ford 494} & \footnotesize{2.19} & \footnotesize{2.3$\pm$0.6} & \footnotesize{0.05$\pm$0.01} & \footnotesize{5$\pm$3} & \footnotesize{1.70/3} & \footnotesize{0.636} & \footnotesize{73} & \footnotesize{117}\\
\footnotesize{r3-7} & \footnotesize{00:43:21.063} & \footnotesize{+41:17:50.44} & \footnotesize{Ford 209} & \footnotesize{2.31} & \footnotesize{2.0$\pm$0.2} & \nodata & \footnotesize{0.6} & \footnotesize{13.44/19} & \footnotesize{0.815} & \footnotesize{224} & \footnotesize{4.2}\\
\enddata
\label{spectab}
\end{deluxetable}
\end{landscape}

\begin{deluxetable}{cccccc}
\footnotesize
\tablewidth{4in}
\tablecaption{PNe Narrow Band Fluxes in units of 10$^{-15}$ erg cm$^{-2}$ s$^{-1}$ }
\tableheadfrac{0.01}
\tablehead{
\colhead{{\footnotesize{PN}}} &
\colhead{\footnotesize{{H$\alpha$}}} &
\colhead{\footnotesize{[S II]}} &
\colhead{\footnotesize{[O III]}}&
\colhead{\footnotesize{[S II]/H$\alpha$\tablenotemark{a}}}&
\colhead{\footnotesize{[O III]/H$\alpha$\tablenotemark{b}}}
}
\tablenotetext{a}{A [S II]/H$\alpha$ ratio greater than 0.4 is typical for shock heated gas.  Lower ratios are typical of photo-ionized gas.}
\tablenotetext{b}{[O III]/H$\alpha$ ratios are corrected for absorption assuming $A_V = N_H/1.79 \times 10^{21}$ \citep{predehl1995}, and the standard extinction law.  Where our fits measured $N_H>2\times 10^{21}$ cm$^{-2}$, $N_H=2\times 10^{21}$ cm$^{-2}$ was assumed.}
\startdata
Ford 322 & 3.4 & 3.1 & 3.4 & 0.9 & 1.3\\
Ford 21 & 1.1 & 1.5 & 11.4 & 1.4 & 13.4\\
Ford 13 & 5.9 & 6.5 & 9.1 & 1.1 & 1.9\\
Ford 201 & 0.9 & 0.6 & 3.3 & 0.7 & 4.7\\
Ford 494 & $<$0.1 & $<$0.1 & 1.0 & \nodata & $>$10\\
Ford 209 & 3.4 & 0.7 & 11.2 & 0.2 & 3.6\\
\enddata
\label{nbtab}
\end{deluxetable}
\newpage
\clearpage

\end{document}